\documentclass[aps,pre,twocolumn,reprint,longbibliography,titlepage,showpacs,showkeys,amsfonts,amssymb,amsmath]{revtex4-1}
\usepackage{hyperref}
\usepackage{graphicx}
\usepackage{epstopdf}
\usepackage{adjustbox}
\usepackage{color}
\usepackage{bm}
\usepackage{pdfsync}
\usepackage{url}
\usepackage{color}
\newcommand{\bse}{\begin{subequations}}
\newcommand{\ese}{\end{subequations}}
\newcommand{\be}{\begin{equation}}
\newcommand{\ee}{\end{equation}}
\newcommand{\bea}{\begin{eqnarray}}
\newcommand{\eea}{\end{eqnarray}}
\newcommand{\kb}{k_{_{\mathbf{B}}}}
\newcommand{\Hl}{\widehat{H}_{_{\mathrm{L}}}}
\newcommand{\ql}{\widehat{q}_{_{\mathrm{L}}}}

\newcommand{\chiq}{\chi_{_{\mathbf{q}}}\!}
\newcommand{\chiqs}{\widehat{\chi}_{_{\mathbf{q}}}\!}
\newcommand{\chiqp}{\dot{\chi}_{_{\mathbf{q}}}\!}
\newcommand{\chiv}{\chi_{_{\mathbf{v}}}\!}
\newcommand{\chivs}{\widehat{\chi}_{_{\mathbf{v}}}\!}
\newcommand{\chivp}{\dot{\chi}_{_{\mathbf{v}}}\!}

\newcommand{\qzero}{q_{_{\mathbf{0}}}}
\newcommand{\vzero}{v_{_{\mathbf{0}}}}

\newcommand{\nun}{\nu_{_{n}}}
\newcommand{\lambdaone}{\lambda_{_{\mathbf{1}}}}
\newcommand{\lambdatwo}{\lambda_{_{\mathbf{2}}}}

\newcommand{\phiv}{\varphi_{_{\!\mathbf{v}}}}
\newcommand{\phiq}{\varphi_{_{\!\mathbf{q}}}}

\newcommand{\BR}{\mathcal{B}\big(R(t)\big)}

\allowdisplaybreaks

\begin{document}
\title{Susceptibility of quasiclassical Brownian motion in harmonic nonlinear potentials}
\author{Pedro J. Colmenares}
\email{gochocol@gmail.com}
\affiliation{Departamento de Qu\'{\i}mica. Universidad de Los Andes. M\'erida 5101, Venezuela}
\thanks{Corresponding Author}
\thanks{Accepted for publication Phys. Rev. E {\bf 102} 06210 (2020). DOI: 10.1103/PhysRevE.102.062102}


\begin{abstract}
This work sets the exact equations for the quasiclassical response function and susceptibility of a Brownian particle immersed in a bath of quantum harmonic oscillators driving by nonlinear harmonic potentials. A delta force perturbation gives rise to a response whose susceptibility is the combination of a linear term, own of the harmonic oscillator, plus a nonlinear one involving an integral \textcolor{black}{equation. It is provided a recursion method to find its solutions based on  functional equations in the Banach space.} The ODE for the response function is a highly nonlinear damped non-autonomous Duffing equation for which the aforementioned method is used to get its solution.
\end{abstract}

 \pacs{05.30.Ch; 05.30.?d; 05.40.Jc; 02.50.Ey}
\keywords{Quantum statistical mechanics, Brownian motion, Stochastic processes, Fokker Planck equation.}
\maketitle
\section{Introduction}

The response of a system to an external disturbance depends on the characteristics of the dynamics and of the applied force. Linear response theory allows one to determine the resonant frequencies of the system with those of the disturbance. Therefore, it is expected that for nonlinear dynamics, the resonant frequencies different from those of the external field will appear as nonlinear terms depending on the magnitude of the force. Other properties, such as the correlation function of the physical variables \cite{GrabertWeissTalkner}, as well as transport coefficients \cite{PJPRE2}, can be determined once the susceptibility has been calculated.

The statistical analysis of fluctuations in the Markovian classical  Langevin equation with non-linear harmonic potentials is a well-studied problem. The Fokker-Planck equation associated with the dynamics of the system is determined with stochastic techniques through the Ito formula \cite{Gardiner1}. Once the eigenvalues of the FPE operator are determined as it is done in Risken's book \cite{Risken}, other properties such  as the Kramers escapes rate and diffusion coefficients, to cite a few, are estimated. This has been done for the normal and metastable (inverted) double-well potential and the asymmetric one, too. A review by Landa {\it et al} \cite{LandaMcClintock} provides a pedagogical presentation of Kramer's formula for the mean first passage time of the Brownian particle leaving any of the bistable wells as a function of its initial position. None of these works attack the calculation of the susceptibility induced by nonlinear potentials. In view of this, it will amenable to extend these well known works to determine their associated nonlinear susceptibilities. It is important that the theory to be developed should reduce to the known non trivial result of the harmonic oscillator (HO) already estimated by Grabert {\it et al} \cite{GrabertWeissTalkner}.   

There is an extensive literature about the theory of nonlinear response. In general, they are suited to the problem at hand. For instance, Diezemann \cite{Diezemann} searched the susceptibility in the Markov approximation of the dielectric relaxation in the reorientational motion of molecules by expanding the probability density function (PDF) in terms of the external field. For quantum systems, Grond {\it et. al.} \cite{Grond} applied a general linear-response many-body theory to trapped interacting bosonic systems and Kryvohuz {\it et. al.} for chemical reactions \cite{Kryvohuz}. Corberi {\it et al.} considered the two-time correlation functions by studying their relation with the  second-order susceptibility in ferromagnetic substances \cite{Corberi} while Bouchaud {\it et al.} made it for glassy ones \cite{Bouchaud}. \textcolor{black}{Marini {\it et al.} \cite{Marini} addressed an extensive review of response theory in stochastic and Fokker-Planck models (see the references there in).}

The objective of this work is to formally characterize with stochastic calculus and standard statistical mechanics tools, the solution, response function and susceptibility associated to the quasiclassical Langevin equation (QCLE) \cite{ PJPRE} with a general nonlinear harmonic potential bias in the configuration space. The equilibrium system is disturbed with an external delta force such that a linear response term accompanies a nonlinear one. The nonlinearity shows up as an integral term depending on the property \textcolor{black}{to be determined} and the dissipation throughout the quantum noise correlation function. It is in some respect, a partial complement to the analysis gained in the dynamical description of the damped HO by Graber {\it et al.}  \cite{GrabertWeissTalkner} to the reign of nonlinear harmonic oscillators. The generalized potential used in this article predicts the results of the HO as a particular case. Therefore, any extra terms are due to the non-linearity of the potential.

The paper is structured as follows. The relevant equations are presented in Sec. \ref{sec1} divided in three subsections. Section \ref{sec1a} covers the system dynamical equation and its solution in particular, the  probability distribution function (pdf) and  associated moments and standard deviation of a Brownian particle submerged in a quantum harmonic oscillator bath under the effect of a general external nonlinear potential. Subsequently in Sec. \ref{sec1b},  is shown the derivation of the response function differential equation and  susceptibility using the solution of the QCLE. Finally, Sec. \ref{sec1c} shows that the response ODE is in fact a special kind of the Duffing equation. It is provided its solution with the method used in Sec \ref{sec1b}. The paper ends in Sec. \ref{sec2} with some final remarks.

\section{Relevant equations}
\label{sec1}
\subsection{Solution of the QCLE}
\label{sec1a}

The dynamics of a Brownian particle immersed in a reservoir of quantum harmonic oscillators at the temperature $T$ with a frequency dependent damping spectra in a position-time dependent potential is of a great interest for its application in quantum systems. This has been done through the path integral approach. Two approaches have been developed: either Caldeira {\it et al.}'s \cite{CaldeiraLeggett} or Grabert {\it et al.}'s \cite{GrabertSchrammIngold}. Their difference lies in the way the initial total density matrix is handled. In the former, the total density is factorized such that there is not correlation between the initial states of the particle and the bath while in the latter is incorporated. 
This leads that Caldeira's equation of motion is not homogeneous leaving an extra term due to the preparation procedure.  Grabert formalism is chosen  in this research. In it, is supposed that the field only affects the particle Hamiltonian leaving unaltered  the bath and that in addition, the time dependent friction term obeys Kubo's second fluctuation-dissipation (SFD) theorem. It is an approximated theory in view of the molecular dynamics results of Daldrop {\it et al} \cite{Daldrop}. In it, a solute-water molecules system in a harmonic oscillator field shows that the friction coefficient depends on the strength of the field. This physical fact requires one to modify the bath Hamiltonian to include it. Recently, Lis\'y and T\'othova \cite{Lisy} did that and showed that in the framework of the generalized Langevin equation (GLE), this extra term in the Hamiltonian modifies the SFD and therefore, the time dependent friction  will depend upon this strength. A similar result for a time dependent field was previously found by Olivares and Colmenares \cite{WorPJPhysA} . The incorporation of the field strength will require the reformulation of the functional integral approach that is out of scope in this research.

Using the premises of Grabert {\it et al} \cite{GrabertSchrammIngold}, the resulting equation of motion is the  quasiclassical GLE with a noise correlation function provided by the theory itself and defined later; that is

\be
\ddot{q}(t)=-\int_{0}^{t}dy\,\Gamma(t-y)\dot{q}(y)-\frac{\partial (V(q,t)}{\partial t} +\xi(t),
\ee
where $q$ is the position, $V(q,t)$ the external potential and the friction kernel $\Gamma(t)$ is the time dependent damping.

Assuming a frequency independent friction $\Gamma(t)=2\gamma\,\delta(t)$ with $\gamma$ being the static friction coefficient and a time independent general nonlinear potential $V(q)=\eta \,q^{2}/2+\alpha\, q^{4}/4+\epsilon\, q$,
encompassing  parabolic ($\eta=1;\,\alpha=\epsilon=0$), bistable ($\eta\,=\,1;\,\,\alpha\,>\,0; \,\, \epsilon=0$) and asymmetric bistable ($\eta\,=-1\,\,\alpha\,>\,0;\,\, \epsilon >0$), the resulting dynamics is controlled by the following Ohmic quasiclassical Langevin equation (QCLE).

\be
\ddot{q}(t)=-\gamma\,\dot{q}(t)-\eta\,q(t)-\alpha\, q(t)^{3}-\epsilon+\xi(t).\\
\label{QLE}
\ee
The quantum noise $\xi(t)$ depends on the initial position of the Brownian particle and the initial \textcolor{black}{momentum and position} of  the bath oscillators \cite{Ingold}. It is Gaussian with zero mean and with a two--time correlation function 
given by
\bea
\left<\xi(t)\,\xi(s)\right>=&-&\left(\frac{\gamma\, T}{2}\right)\nu\sinh^{-2}\left[\frac{1}{2}\nu(t-s)\right]\nonumber\\
&+&\,i\frac{2\,\pi\,\gamma}{\nu}\dot{\delta}(t-s),\label{corxi}
\eea
where $\nu$ is the reduced Matsubara frequency \cite{PJPRE}.

Unlike the classical Markovian Langevin equation, the noise has a colored spectrum and is also correlated with the initial position $\qzero$ because of the preparation procedure. The latter is \cite{SchrammJungGrabert}:

\bea
\left<\xi(t)\qzero\right>=-2\,\gamma\,T\sum_{n=1}^{\infty}\frac{\nun}{(\nun\!+\!\lambdaone)(\nun\!+\!\lambdatwo)}\mbox{e}^{-\nun t},
\label{initcor}
\eea
where $\nun=n\nu$ and frequencies $ \lambda_{_{\mathbf{1,2}}}=(\gamma\pm(\gamma^{2}- 4)^{1/2}\,)/2$. 

Equation (\ref{QLE}) is obtained from the physical QCLE by scaling energy, position and time by the parameters $\kappa\,q_{m}^{2}$, $q_{m}$ and $(M/\kappa)^{1/2}$, respectively, where $M$ is the mass of the particle, $\kappa$ is the stiffness of the harmonic term of the potential and $q_{m}$ is chosen arbitrarily. For the bistable and asymmetric bistable potentials could be the minimum located at the right, while for the HO is of free choice except $q_{m}=0$.  $T=\kb \widetilde{T}(\kappa \,q_{m}^{2})^{-1}$  where $\widetilde{T}$ is the actual temperature in kelvin and $\kb$ is the Boltzmann constant.

The QCLE differs from the Markovian classical LE in which the noise correlation is still Gaussian but delta correlated. The noise average of the HO QCLE is indistinguishable from the classical. Thus, any calculations made on it will give the classical result. The QCLE extends to the nonlinear reign of the HO Markovian \textcolor{black}{classical} Langevin equation.

The statistical characterization of the dynamics driven by Eq. (\ref{QLE}) is based on the adaptation of the method developed in \cite{PJPRE} for the HO to find the PDF and its standard deviation.

The Laplace transformation of the QCLE is:
\bea
\ql(s)&=&\chiqs(s)\,\qzero+\chivs(s)\,\vzero-\frac{\epsilon}{s}\,\chivs(s)\nonumber\\
&+&\chivs(s)\bigg(\widehat{\xi}(s)-\alpha\,\Hl(s)\bigg),\label{lapeq}
\eea
where $\qzero$ and $\vzero$ are the initial position and velocity,
\bea
\chiqs(s)&=&(s+\gamma)\chivs(s),\\
 \chivs(s)&=&\frac{1}{s^{2}+\gamma\,s+\eta}\label{chivs},
\eea
 and function $\Hl(s)=\mathcal{L}\big\{q^{3}(t)\big\}$ obeys the identity
\bea
\mathcal{L}\big\{q^{3}(t)\big\}&=&\Bigg(\mathcal{L} \bigg\{\big(q^{3}\star q^{3}\star q^{3}\big)(t)\bigg\}  \Bigg)^{1/3},
\eea
with the triple convolution given by 
\be
\big(q^{3}\star q^{3}\star q^{3}\big)(t)=\int_{0}^{t}\!\!dy\,q^{3}(t-y)\int_{0}^{y}\!\!dx\,q^{3}(y-x)\,q^{3}(y).
\ee

After applying the inverse transformation it is found that
\bea
q(t)&=&\bar{q}(t) +\phiq(t).\\
v(t)&=&\bar{v}(t)+\phiv(t),\\
\bar{q}(t)&=&\textcolor{black}{\chiv(t)\,\vzero+\chiq(t)\,\qzero}\nonumber\\
&\textcolor{black}{+}&\textcolor{black}{\int_{0}^{t}dy\bigg[\epsilon\chiv(y)-\alpha\chiv(t-y)\,H(y)\bigg]},\label{qbar}\\
\bar{v}(t)&=&\textcolor{black}{\chivp(t)\,\vzero+\chiqp(t)\,\qzero+\epsilon\,\chiv(t)}\nonumber\\
&-&\textcolor{black}{\int_{0}^{t}dy\,\alpha\,\chivp(t-y)\,H(y)}\label{vbar},
\eea
where the noises $\phiq(t)$ and $\phiv(t)$ are defined by
\bea
\phiq(t)&=&\int_{0}^{t}dy\,\chiv(t-y)\,\xi(y),\label{phiq}\\
\phiv(t)&=&\int_{0}^{t}dy\,\chivp(t-y)\,\xi(y)\label{phiv}.
\eea
The \textcolor{black}{over bar} on $q(t)$ and $v(t)$ denotes their average values and the susceptibilities $\chiq(t)$ and $\chiv(t)$ read as
\bea
\chiq(t)&=&\mbox{e}^{-\gamma\,t/2}\Bigg[\cosh\bigg[\frac{\omega_{0}}{2}t\bigg]+\frac{\gamma}{\omega_{0}}\sinh\bigg[\frac{\omega_{0}}{2}t\bigg]\Bigg],\\
\chiv(t)&=&\frac{2}{\omega_{0}}\mbox{e}^{-\gamma\,t/2}\sinh\bigg[\frac{\omega_{0}}{2}t\bigg],\label{chivt}
\eea
with $\omega_{0}=\sqrt{\gamma^{2}-4\,\eta}$. The presence of the nonlinear term in the potential shows up through $\Hl(s)$. Therefore, the integral equation in the Laplace space given by Eq. (\ref{lapeq}) has to be first numerically solved  in order to get all dynamical properties related to $q(t)$. This partial result is not a simple numerical problem of computing. Apart from that, it is own of the mathematical formulation.

Since the equations for $q(t)$ and $v(t)$ have the same structure as those used in Ref. \cite{PJPRE}, the conditional distribution $p(q,t|\qzero)$ is Gaussian with mean $G(t)$ and standard deviation $\sigma^{2}(t)$ given by  
\bea
G(t)&=&\textcolor{black}{\chiq(t)\,\qzero\!\!+\!\!\int_{0}^{t}\!\!\!dy\bigg[\epsilon\chiv(y)-\alpha\chiv(t\!-\!y)\,H(y)\bigg]},\\
\textcolor{black}{\sigma^{2}(t)}&=&T\,\chiv^{2}(t)+2\int_{0}^{t}dy\Bigg[\int_{0}^{y}dz\,\big<\phiv(z)\phiv(y)\big>\nonumber\\
&+&\chiq(y)\,\big<\phiv(y)\qzero\big>\Bigg]\label{sigma},
\eea
respectively. The $\eta$ dependence on $G(t)$ is controlled by the susceptibilities $\chiq(t)$ and $\chiv(t)$. Similarly, the amplitude of the non-linear term affects the standard deviation thorough $\chiv^{2}(t)$. 

It is worth mentioning that the preceding procedure can be applied to the generalized QCLE. Important to keep in mind is that $\chivs(s)$, Eq. (\ref{chivs}), will not depend on $\gamma$ but on the Laplace transform of the kernel, that is, $\widehat{\Gamma}(s)$. In order to get a closed equation, is imperative to define first the functional form of the time-dependent kernel \cite{WorPJPhysA}.

\subsection{Response Function ODE and Susceptibility}
\label{sec1b}

According to Ehrenfest's theorem, Eq. (\ref{QLE}) reads:
\bea
F(t)&=&\Big<\ddot{q}(t)\Big>+\gamma\,\Big<\dot{q}(t)\Big>+
\Big(3\,\alpha\,\textcolor{black}{\sigma^{2}(t)}+\eta\Big)\,\Big<q(t)\Big>\nonumber\\
&+&\alpha\,\Big<q(t)\Big>^{3}+\epsilon,
\label{QLEnew}
\eea
where the external driving force $F(t)$ was added . The third moment of $q(t)$ was written as $\big<q^{3}(t)\big>=\big(\big<q(t)\big>^{3}+3\,\big<q(t)\big>\,\textcolor{black}{\sigma^{2}(t)}\big)$ with \textcolor{black}{$\sigma^{2}(t)$} given by Eq. (\ref{sigma}).

The response function $R(t)$ of the system as the result of the perturbation driving by $F(t)$ is defined by $\big<q(t)\big>=\int_{-\infty}^{\infty}dy\,R(t-y)\,F(y)$. Then for a driving force $F(t)=F_{0}\,\delta(t)$, we obtain the following ODE for $R(t)$
\bea
\delta(t)&=&\ddot{R}(t)+\gamma\,\dot{R}(t)+\Big(3\,\alpha\,\sigma^{2}(t)+\eta\Big)R(t)\nonumber\\
&+&\alpha\,F_{0}^{2}\,R^{3}(t)+\frac{\epsilon}{F_{0}},\label{response}
\eea
with initial conditions $R(0)=0$ and $\dot{R}(0)=1.$

This equation for finite values of $\alpha$ and $\epsilon$ gives rise to a non linear response due to the dependence on $F_{0}$. For the parabolic potential, the perturbation excites a response of the same frequency \cite{Reichl}.   

The susceptibility $\chi(\omega)$ is defined as the Fourier transform (FT) of the response function, that is,
$\chi(\omega)=\int_{0}^{\infty}dt\,R(t)\mbox{e}^{\imath\,\omega\,t}$. Recalling the convolution identities
\bea
\mathcal{F}\Big\{R(t)\,\sigma(t)\Big\}&=&\int_{-\infty}^{\infty}dy\,\chi(\omega-y)\,\widehat{\textcolor{black}{\sigma^{2}}}_{_{\!\mathrm{F}}}(y),\\
\mathcal{F}\Big\{R^{3}(t)\Big\}&=&\int_{-\infty}^{\infty}dy\,\chi(\omega-y)\nonumber\\
&\times&\int_{-\infty}^{\infty}dz\,\chi(z)\,\chi(y-z),
\eea
then, the FT of the response function's ODE, Eq. (\ref{response}), gives the susceptibility,
\bea
\chi(\omega)&=&=\phi(\omega)+\psi\big(\omega\big),\label{susc}\\
\phi(\omega)&=&\widetilde{\chi}(\omega)\bigg(1\textcolor{black}{-}\frac{\epsilon}{F_{0}}\delta(\omega)\Bigg),\\
\psi\big(\omega,[\chi]\big)&=&\widetilde{\chi}(\omega)\int_{-\infty}^{\infty}d\omega^{\prime}\,\chi(\omega-\omega^{\prime})\Bigg[3\,\widehat{\textcolor{black}{\sigma^{2}}}_{_{\!\mathrm{F}}}(\omega^{\prime})\nonumber\\
&\textcolor{black}{-}&\textcolor{black}{\alpha}\,F_{0}^{2}\int_{-\infty}^{\infty}d\omega^{\prime\prime}\,\chi(\omega^{\prime\prime})\,\chi(\omega^{\prime}-\omega^{\prime\prime})\Bigg],\label{Banachchi}
\eea
where $\widetilde{\chi}(\omega)=(\eta-\omega^{2}-\imath\,\gamma\,\omega)^{-1}$. It is a rather complicated equation for non-linear potentials that involves solving a Fredholm integral equation of the second kind. However, for the HO ($\eta=1\,;\,\alpha=\epsilon=0$) $\chi(\omega)= \widetilde{\chi}(\omega)$ agrees with that of the reported in the literature \cite{GrabertWeissTalkner,Reichl}.

The proposed solution of this integral equation uses a method developed by Daftardar-Gejji {\it et al.} \cite{Daftardar}. Its application to a variety of integral equations including ODEs, are of a high order of accuracy \cite{Daftardar,Jawary}. It has been chosen because of its simplicity. In order to show the scope of the algorithm, it will be fully described next. 

Considering $\psi\big(\omega,[\cdot]\big)$ as the Banach operator $\mathcal{\textcolor{black}{B}}$ in the Banach space $\mathcal{B}\rightarrow \mathcal{B}$ and $f(\omega)=\phi(\omega)$, then Eq. (\ref{susc}) can be written as
\be
\chi(\omega)=f(\omega)+\mathcal{B}\big(\chi(\omega)\big).\label{chiBanach}
\ee
Let the solution be given by
\be
\chi(\omega)=\sum_{i=0}^{\infty}\chi_{i}(\omega),
\ee
and decomposw the operator $\mathcal{B}\big(\chi(\omega)\big)$ according to
\bea
\mathcal{B}\Bigg(\sum_{i=0}^{\infty}\chi_{i}(\omega)\Bigg) &=&\mathcal{B}\big(\chi_{_{0}}(\omega)\big)+\sum_{i=1}^{\infty}\Bigg[\mathcal{B}\Bigg(\sum_{j=0}^{i}\chi_{j}(\omega)\Bigg) \nonumber\\
&-&\mathcal{B}\Bigg(\sum_{j=0}^{i-1}\chi_{j}(\omega)\Bigg)\Bigg],
\eea
which is equivalent to
\bea
\sum_{i=0}^{\infty}\chi_{i}(\omega) &=& f(\omega)+\mathcal{B}\big(\chi_{_{0}}(\omega)\big)\\
&+&\sum_{i=1}^{\infty}\!\Bigg[\mathcal{B}\Bigg(\!\sum_{j=0}^{i}\chi_{j}(\omega)\Bigg) \! -\! \mathcal{B}\Bigg(\!\sum_{j=0}^{i-1}\chi_{j}(\omega)\Bigg)\!\Bigg].
\eea
Defining the following recursion relation:
\bea
\chi_{_{0}}(\omega)&\textcolor{black}{=}& f(\omega),\\
\mathcal{\chi_{_{1}}}(\omega)&=&
\mathcal{B}\bigg(\chi_{_{0}}(t)\bigg),\\
\chi_{m+1}(\omega)&=&\mathcal{\textcolor{black}{B}}\bigg(\chi_{_{0}}(\omega)+\cdots+\chi_{m}(\omega)\bigg)\nonumber\\
&-&\mathcal{\mathcal{B}}\bigg(\!\!\chi_{_{0}}(\omega)+\cdots+\chi_{m-1}(\omega)\!\!\bigg);\,\,m\geqslant 1,
\eea
then
\be
\chi(t)=f(\omega)+\sum_{i=1}^{\infty}\chi_{i}(\omega).\label{sol}
\ee
Thus the solution is recursively calculated through each of the $\chi_{i}(\omega)$. The convergence of the summation to be one of the set of possible solutions is guaranteed  in view of the Banach fixed-point theorem \cite{Daftardar}. Since $\chi(\omega)$ is complex, Eq. (\ref{sol}) can be written as a system of equations for its real and imaginary parts. In particular, the position correlation function is written as an integral of its imaginary part with poles located  in the lower half plane with values given by the roots of the denominator of the imaginary part of $\widetilde{\chi}(\omega)$ as in the HO \cite{GrabertWeissTalkner}.

Apart from $\chi_{_{0}}(\omega)$, functions of superior order requires one to know $\widehat{\sigma^{2}}_{_{\!\mathrm{F}}}(\omega^{\prime})$ which involves the FT of the multidimensional integral given by Eq. (\ref{sigma}). Using the method described in Ref. [2] allows one to write down the noise correlation in terms of the numerical position correlation function. Once the standard deviation and its FT are calculated then it can be fitted as a given function and subsequently used in the integral depicted in Eq. (\ref{Banachchi}). In general, the different terms  will have among others, large polynomials in their constitutive equations because the \textcolor{black}{zero} order term will appear in each of the $\chi_{i}(\omega)$. As a collateral result, once the noise correlation is calculated then the diffusion constant of the system can also be determined as is exemplified in \cite{PJPRE2} for the parabolic potential.

Having known $\chi(\omega)$ then $R(t)$ is given by its inverse, that is,

\be
\textcolor{black}{R(t)=\frac{1}{2\,\pi}\,\int_{-\infty}^{\infty}d\omega\,\chi(\omega)\,\mbox{e}^{-\imath\,\omega\,t}.\label{def}}
\ee

This is the usual method found in textbooks to get the response function. In the next section the application of the Banach operator method to solve the response function ODE will be shown.

\section{Response function ODE and Duffing equation}
\label{sec1c}

Equation (\ref{response}) is a highly non linear damped non autonomus Duffing equation (DE) for which the literature about solution methods and application for specific systems is very rich. A good review of the application to control systems is found in  Chap. 5 of Kovacic and Brennan's book \cite{Kovacic} and approximated methods of solutions in Nayfeh's \cite{Nayfeh}.  The methods are primarily based on perturbation techniques of some small parameter \cite{Nayfeh}. Although they have been formally deducted, its algebra is cumbersome and lengthy. Instead, the operator method discussed above will be apply.

The conversion of the ODE into the functional form such as given by Eq. (\ref{chiBanach}) is simple. To do that, let $L_{tt}(\cdot)=\partial^{2}(\cdot)/\partial t^{2}$, $L_{t}(\cdot)=\partial(\cdot)/\partial t$, $L_{tt}^{-1}\,(\cdot)=\int_{0}^{t}dy\int_{0}^{y}ds(\cdot)$ and $L_{t}^{-1}(\cdot)=\int_{0}^{t}dy\,(\cdot)$. Then, applying the operator $L_{tt}^{-1}(\cdot)$ to the response function's ODE, using the initial conditions and  that $L_{tt}^{-1}(\cdot)=\int_{0}^{t}dy\,(t-y)(\cdot)$, the resulting functional equation is of the type of Eq. (\ref{chiBanach}, with
\bea
f(t)&=&\frac{1}{2}\bigg(4-\frac{\epsilon}{\textcolor{black}{F_{0}}}t\bigg)t,\\
\textcolor{black}{\BR}&\textcolor{black}{=}&\textcolor{black}{-\int_{0}^{t}dy\,\Bigg\{\gamma\,R(y)+(t-y)\bigg[\eta\,R(y)}\nonumber\\
&-&\textcolor{black}{\alpha\bigg( F_{0}^{2}\,R^{3}(y)\!+\!3\,R(y)\,\sigma^{2}(y)\bigg)\bigg]\Bigg\}},\label{Bsigma}
\eea
with Eq. (\ref{Bsigma}) defining the operator $\mathcal{B}$ of the system. The functional equation is  a Volterra integral equation type such that the method mentioned above can be used to find its solution. Each of the $R_{i}(t)$ are calculated by the recursion formula shown above. \textcolor{black}{The method of analysis described in Ref. \cite{Kovacic} could be used to characterize the dynamical aspects of the solution due to the nonlinearity.}

Unfortunately, as in the solution of the susceptibility equation, we cannot analytically go any further but numerically, because the Banach operators requires the standard deviation, which in turn must also be determined numerically.

\section{Final Remarks}
\label{sec2}

The calculation of the susceptibility associated to the QCLE with nonlinear potentials using standard tools of stochastic equations and statistical mechanics leads to an integral equation for this property. It has such a complexity that numerical methods have to be used to solve it. However, the proposed recursive method of Banach operators reduces in a large extension the calculations bypassing any iterative technique in the solution. In any case, the incorporation of nonlinear potentials and the tools already mentioned do not allow getting a basic physical insight about the susceptibility just as that found for the pure harmonic oscillator. Doing numerical calculations can only elucidate it. This drawback opens the doors to find alternative routes to tackle this problem in order to obtain at least simpler equations. In any case, the equations are exact without any assumptions made in their derivations.

The Banach operator method for functional equations is physically sound. It shows how any 
function can be expanded up to a specified order so that additional calculations whose contributions are negligible can be prevented. The maximal error remainder test \cite{Jawary} can be used to stop the numerical procedure. This is guaranteed by the Banach fixed-point theorem.The derivation of the formulas was possible because the PDF or Green function of the dynamics is completely known.  In cases where 
the PDF is not available, the theory of Diezemann \cite{Diezemann} properly 
adapted to the problem at hand, would be the route to find the different orders of the response function.

The problem of classical Markovian systems is easier to handle because the noise is delta correlated. In particular the standard deviation is analytical \cite{PJPRE2}. 

A previous work \cite{PJPRE2} mentions the  class of systems potentially can be used with this formalism.
\section*{Acknowledgments}
The author is grateful to \textcolor{black}{the referees} for helpful suggestions.

%

\end{document}